%% file: main.tex
\documentclass[twocolumn]{aastex701}
\usepackage{amsmath}
\usepackage{multirow}

\newcommand{\sample}{37}
\newcommand{\dmhost}{\ensuremath{\mathrm{DM_{host}}}}
\newcommand{\mstar}{\ensuremath{M_\star}}
\newcommand{\dmm}{\dmhost--\mstar}
\newcommand{\logm}{\ensuremath{\log_{10}(M_\star/M_\odot)}}

\shorttitle{The $\mathrm{DM}_{\mathrm{host}}$--$M_\star$ Relation}

\shortauthors{Athukoralalage et al.}

\begin{document}

\title{The Local Origin of a Host Dispersion Measure–Stellar Mass Relation in Fast Radio Bursts}

\correspondingauthor{Wasundara Athukoralalage}
\email{wasundara.athukoralalage@cfa.harvard.edu}

\author[orcid=0000-0002-2866-6416, gname=Wasundara,sname=Athukoralalage]{Wasundara Athukoralalage}
\affiliation{Center for Astrophysics \textbar{} Harvard \& Smithsonian, 60 Garden Street, Cambridge, MA 02138-1516, USA}
\email[]{wasundara.athukoralalage@cfa.harvard.edu}  

\author[orcid=0000-0002-7587-6352]{Liam Connor} 
\affiliation{Center for Astrophysics \textbar{} Harvard \& Smithsonian, 60 Garden Street, Cambridge, MA 02138-1516, USA}
\email[]{liam.connor@cfa.harvard.edu}

\author[0000-0003-0526-2248]{Peter K.~Blanchard}
\affiliation{Center for Astrophysics \textbar{} Harvard \& Smithsonian, 60 Garden Street, Cambridge, MA 02138-1516, USA}
\affiliation{The NSF AI Institute for Artificial Intelligence and Fundamental Interactions, USA}
\email[]{peter.blanchard@cfa.harvard.edu}

\author[orcid=0000-0002-9392-9681]{Edo Berger} 
\affiliation{Center for Astrophysics \textbar{} Harvard \& Smithsonian, 60 Garden Street, Cambridge, MA 02138-1516, USA}
\affiliation{The NSF AI Institute for Artificial Intelligence and Fundamental Interactions, USA}
\email[]{eberger@cfa.harvard.edu}

\begin{abstract}

The origin of fast radio bursts (FRBs) and the relationship to their host galaxy environments remain open questions. Using a sample of \sample\ FRBs with secure host associations at $z \lesssim 0.25$, and including dwarf, elliptical, and edge-on hosts, we examine the host dispersion measure $(\mathrm{DM}_{\mathrm{host}})$ and its dependence on host stellar mass ($M_\star$). We find a statistically significant anti-correlation between the host dispersion measure and host stellar mass (the \dmm\ relation; Pearson $r \approx -0.5$, $\log_{10}p \approx -2.8$) that extends across the full range of our sample, \logm\ $\approx 8$--$11.4$. We find that the rest-frame Faraday rotation measure ($|\mathrm{RM}|_{\rm host}$) is largest in FRBs that reside in dwarf galaxies, but there is no evidence for a correlation across the full mass range. We show that the \dmm\ trend does not arise from the galaxy-averaged interstellar medium (ISM) nor from the circumgalactic medium (CGM). Instead, we argue that the local ($\lesssim100$\,pc) environment is responsible. The spatial statistics of star formation and metallicity-dependent FRB progenitor channels plausibly lead to denser ionized environments in low-mass galaxies and less nearby gas in massive hosts.

\end{abstract}


\section{Introduction} \label{sec:introduction}

Fast radio bursts (FRBs) are bright and luminous $\mathrm{(\gtrsim 10^{42}\ erg\ s^{-1})}$, millisecond-duration pulses at 0.1--5\,GHz  \citep{lorimer_bright_2007, Cordes_2019, petroff_fast_2022}, whose origin remains a mystery. Their dispersion measures (DMs), which trace the integrated electron column density along the line of sight, greatly exceed the contributions from the Milky Way (MW) and its halo, implying an extragalactic origin. This has now been confirmed by the identification of $\sim\!100$ FRB host galaxies \citep{chatterjee_direct_2017, bhandariCharacterizingFastRadio2022, heintz_host_2020, law_deep_2024, sharma_preferential_2024,connor_gas-rich_2025, gordon_demographics_2023}. While most FRBs have been detected only once, repeat bursts have been observed from a subset of sources \citep{spitler_repeating_2016,the_chimefrb_collaboration_chimefrb_2019,the_chimefrb_collaboration_chimefrb_2023,fonseca_nine_2020}, and it is possible that all FRBs repeat on a wide range of timescales (e.g., \cite{ravi_prevalence_2019, james_modelling_2023}). Despite the availability of galaxy-scale properties, the sources that produce FRBs, and their evolutionary pathways remain elusive, with dozens of proposed models (e.g., \cite{platts_living_2019}). The lack of a clear understanding around FRB origins is due to host galaxies spanning a variety of masses and types, and most localizations being too crude ($\sim\!10$ kpc) to enable an association with a specific stellar population or individual sources within the galaxies. 

The short, energetic, and polarized radio bursts of FRBs require a compact emission region, pointing to an association with black holes or neutron stars \citep{platts_living_2019}, and in particular magnetars, whose large magnetic fields ($10^{13}-10^{15}$\,G) can provide the required energy budget. Magnetars are sufficiently abundant to match the large volumetric rate of FRBs ($\geq10^4$\,Gpc$^{-3}$\,yr$^{-1}$). Although most FRBs arise in star-forming galaxies (including both massive spirals and dwarfs), as expected for the formation of young compact objects in core-collapse supernovae \citep{chatterjee_direct_2017, niu_repeating_2022, sharma_preferential_2024}, they have also been identified in evolved galaxies dominated by old stellar populations \citep{kirsten_repeating_2022,shah_repeating_2025,eftekhari_massive_2025}, potentially pointing to formation channels such as accretion-induced collapse of white dwarfs, or the remnant of a neutron star binary merger. The diversity in galaxy-scale properties precludes a definitive answer to the nature of FRB sources and their formation pathways in current data.

Recent work by \cite{leung_stellar_2025} argued that the relationship between the stellar mass of a galaxy and its host DM is a sensitive probe of baryonic feedback. The authors report an anti-correlation between host galaxy DM and stellar mass for $L^*$ galaxies, showing that their data are inconsistent with hydrodynamical simulations in which feedback is low (e.g. CAMELS/Astrid). Their sample excluded massive elliptical hosts, edge-on spiral galaxies, scattered FRBs, galaxy cluster hosts, and dwarf galaxies to isolate the circumgalactic medium (CGM) contribution. For CGM studies, sources that show indications of high non-CGM DM contributions should be omitted. While the modest host DMs of the high mass end supports their claim of efficient feedback, the CGM alone is not expected to provide enough ionized gas column to explain the observed host DM for $\log M_\star \lesssim 9.5$. 

In this paper, we sample a wide range of nearby localized FRBs ($z \lesssim 0.25$) with secure host associations and find that the host
dispersion measure--stellar mass (\dmm) anti-correlation is likely tracing local ($\lesssim100$\,pc) FRB environments, or a connection between low-mass galaxies and unusually dense star-forming or circumburst regions. We find that this correlation extends to the full range of stellar masses (\logm\ $\approx 8$--$11.4$) and that host DM is a tracer of local environments.

The structure of this paper is as follows. In \S\ref{sec:sample-selection} and \S\ref{sec:data-analysis} we describe the sample selection of FRBs and how their data were compiled. In \S\ref{sec:results}, we present the result of our analysis. We discuss the results in \S\ref{sec:discussion} and outline the main conclusions in \S\ref{sec:conclusions}. Throughout the paper we assume a flat $\Lambda$CDM cosmology based on the Planck 2015 results with \mbox{$H_{0} = 67.8$ km s$^{-1}$ Mpc$^{-1}$} and $\Omega_{m} = 0.309$ \citep{planck2015}. Throughout, we write $\log M_\star$ as shorthand for \logm.

\section{Sample Selection}\label{sec:sample-selection}

We construct a sample of FRBs with secure host associations reported in the literature through August 7, 2026. We impose a redshift cutoff of $z \lesssim 0.25$ to reduce the uncertainty in cosmic DM ($\mathrm{DM}_{\mathrm{cos}}$) associated with distant FRBs. It is difficult to identify low-mass hosts and constrain DM from intergalactic medium (IGM) beyond $z \approx 0.25$. We construct the sample using the following criteria:

\begin{enumerate}

\item We require a two-dimensional interferometric localization of around (sub)arcsecond precision, for a secure host association.
This excludes FRBs localized only along the single 66\,km CHIME--KKO baseline \citep{collaboration_catalog_2025}, whose $2''\times\sim60''$
(1$\sigma$) error ellipses span tens to hundreds of kpc at these redshifts, allowing only probabilistic associations. CHIME sources with two-dimensional
localizations remain in the sample (FRB\,20240209A, \citealp{shah_repeating_2025}; FRB\,20250316A, \citealp{collaborationFRB20250316ABrilliant2025};
and FRB\,20210603A, \citealp{cassanelliFastRadioBurst2024}).

\item We include dwarf and elliptical FRB hosts, in contrast to \citet{leung_stellar_2025}. This adds FRB\,20240209A \citep{eftekhari_massive_2025, shah_repeating_2025}, the only unambiguously elliptical host to date. We include six dwarf ($\log_{10}(M_\star/M_\odot) \leq 9$) hosts: FRB\,20121102A
\citep{spitler_repeating_2016, tendulkar_host_2017}, FRB\,20190417A \citep{moroianu_milliarcsecond_2025}, FRB\,20190520B \citep{niu_repeating_2022}, FRB\,20210117A \citep{gordon_demographics_2023}, FRB\,20230708A \citep{muller26}, and FRB\,20240114A \citep{bhardwaj_hyperactive_2025}. We also choose to include edge-on disk galaxies and scattered FRBs, as we are concerned with total $\mathrm{DM}_{\mathrm{host}}$, which were not used in \citet{leung_stellar_2025} due to their focus on the CGM.

\item We exclude host galaxies that are part of galaxy clusters (including FRB\,20190303A) because the intracluster medium can dominate $\mathrm{DM}_{\mathrm{host}}$.

\item Host stellar mass is ambiguous for repeating source FRB\,20200120E, which resides in a globular cluster (GC) of M81. It remains the only definitive GC FRB, but it is also one of the few sources that \textit{could} have been identified as belonging to a GC. This is because its high repetition rate allowed for VLBI localization and it is the most nearby extragalactic FRB to date. We assign its host stellar mass to that of M81, because if FRB\,20200120E were at the redshift of a typical source in our sample or did not repeat, it would only be associated with the parent galaxy. This choice has implications for the interpretation of our results, which we address in Section~\ref{sec:discussion}.
\end{enumerate}

Applying these criteria results in a sample of \sample\ FRBs with published stellar masses. Table~\ref{tab:frb-sample} shows the sample including the $\mathrm{DM}_{\mathrm{obs}}$ and published host stellar mass for each FRB, together with the rest-frame $\mathrm{DM}_{\mathrm{host}}$ and $|\mathrm{RM}|_{\mathrm{host}}$ values calculated in this work.

\input{frb-sample}


\section{Data Analysis}\label{sec:data-analysis}

The host contribution to the DM budget ($\mathrm{DM}_{\mathrm{host}}$) in the rest frame of the host can be expressed as:

\begin{equation} \label{eq:dmhost}
\mathrm{DM}_{\mathrm{host}} =
(\mathrm{DM}_{\mathrm{obs}}
 - \mathrm{DM}_{\mathrm{MW}}
 - \mathrm{DM}_{\mathrm{cos}}) \times (1+z)
\end{equation}

\noindent $\mathrm{DM}_{\mathrm{obs}}$ is the total observed DM and $\mathrm{DM}_{\mathrm{cos}}$ 
is the cosmic contribution from the IGM and intervening halos. For each FRB, we calculate the $\mathrm{DM}_{\rm MW}$ (contribution from Milky Way) value from the NE2025 model \citep{ne2025} to estimate $\mathrm{DM}_{\rm NE2025}$ (contribution from Milky Way ISM) and the model of \cite{yamasaki_galactic_2020} to estimate $\mathrm{DM}_{\rm YT20}$ (contribution from the Milky Way CGM) using the python package {\tt PyGEDM} \citep{pygedm}\footnote{\url{https://github.com/FRBs/pygedm}}. 
Thus, $\mathrm{DM}_{\rm MW}=\mathrm{DM}_{\rm NE2025}+\mathrm{DM}_{\rm YT20}$. $\mathrm{DM}_{\mathrm{host}}$ can be decomposed further into three components: the host CGM $(\mathrm{DM}_{\mathrm{CGM}})$, the interstellar medium $(\mathrm{DM}_{\mathrm{ISM}})$, and the local/circumburst ionized material $(\mathrm{DM}_{\mathrm{loc}})$.

\begin{equation}
    \rm DM_{host} = DM_{CGM} + DM_{ISM} + DM_{loc}
\end{equation}

\noindent We consider $\rm DM_{loc}$ to be dispersion from an associated H\,\textsc{ii} region, supernova remnant, persistent radio source, or other ionized nebulae in the immediate environment of the source. The mean value of $\rm DM_{CGM}$ and $\rm DM_{ISM}$ are both expected to 
scale with host galaxy mass. The dependence of $\rm DM_{loc}$ on $M_\star$ is unknown, but a relation could exist if the statistics of star formation or FRB stellar populations vary as a function of host stellar mass. We note that a large portion of $\rm \left < DM_{host} \right >$ can come from 
the CGM and ISM, while trends in host galaxy type or mass can come from $\rm DM_{loc}$.

We calculate the probability density of $\mathrm{DM}_{\rm host}$ using,

\begin{equation} \label{eq:dmhost-unc}
\mathcal{P}(\mathrm{DM}_{\rm host}) =\!\!\
\int\limits_{0}^{\rm DM_{max}}
\!\!\!\!\mathcal{P}(\mathrm{DM}_{\rm cos})\,
\mathcal{P}_{\rm MW}\!\left(\mathrm{DM'}_{\rm MW}\right)
\, d\mathrm{DM}_{\rm cos}
\end{equation}

\noindent where $\mathrm{DM'}_{\rm MW}=\mathrm{DM}_{\rm obs}-\mathrm{DM}_{\rm cos} - \rm DM_{host}$ in the integrand, and $\rm DM_{max} = \mathrm{DM}_{\rm obs}-\mathrm{DM}_{\rm host}$.

To evaluate $\mathrm{DM}_{\rm host}$ and its uncertainty, we compute the probability density $\mathcal{P}(\mathrm{DM}_{\rm host})$ defined in Equation~\ref{eq:dmhost-unc}. We use the probability distribution $\mathcal{P}(\mathrm{DM}_{\rm cos} \mid z)$ from \citet{connor_gas-rich_2025}, provided as a grid over cosmic DM and redshift. For each FRB at redshift $z$, we extract the corresponding $\mathcal{P}(\mathrm{DM}_{\rm cos})$ for the nearest tabulated redshift.

We evaluate the $\mathcal{P}(\mathrm{DM}_{\rm host})$ on a uniform grid of $100$ $\mathrm{DM}_{\rm host}$ values that $\in [0, \mathrm{DM}_{\rm obs}]$. For an FRB with an observed dispersion measure $\mathrm{DM}_{\rm obs}$ and Milky Way contribution $\mathrm{DM}_{\rm MW}$, we compute the integral in Equation~\ref{eq:dmhost-unc} numerically as a discrete sum over the $\mathrm{DM}_{\rm cos}$ grid. Here, we model $\mathcal{P}(\mathrm{DM}_{\rm MW})$ as a Gaussian with $\mu = \mathrm{DM}_{\rm MW}$ and a standard deviation $\sigma = 0.2\,\mathrm{DM}_{\rm MW}$ to account for uncertainty in the model,
evaluated at the implied Milky Way contribution $\mathrm{DM}_{\rm obs} - \mathrm{DM}_{\rm cos} - \mathrm{DM}_{\rm host}$. For each $\mathrm{DM}_{\rm host}$ value on the grid, we use only the $\mathrm{DM}_{\rm cos}$ values that satisfy $0 < \mathrm{DM}_{\rm cos} < \mathrm{DM}_{\rm obs} - \mathrm{DM}_{\rm host}$. This prevents us from getting nonphysical negative values for
$\mathrm{DM}_{\rm host}$ for FRBs with high $\mathrm{DM}_{\rm MW}$ values calculated from the model.

We normalize each $\mathcal{P}(\mathrm{DM}_{\rm host})$ to unit area and convert it to the host rest frame by scaling the grid by $(1+z)$. From the
normalized distribution we report the median as our fiducial $\mathrm{DM}_{\rm host}$ and the $16$th and $84$th percentiles as the
asymmetric $1\sigma$ uncertainties. For FRBs whose distribution peaks at $\mathrm{DM}_{\rm host} = 0$, the host contribution is consistent with zero, and we report them in the same way as the rest of the sample and refer to them as having a negligible host DM; they are flagged in Table~\ref{tab:frb-sample} and are not upper limits.

For FRBs with a published rotation measure, we  estimate the Galactic foreground $\mathrm{RM_{MW}}$ along each sight line from the revised Galactic Faraday rotation map \citep{mwrm}, assuming a symmetric uncertainty of 5~rad\,m$^{-2}$ \citep{taylor2023mighteepolarizationearlyscience, mwrm}. As we exclude galaxy cluster hosts, we take $\mathrm{RM_{obs} \approx RM_{MW} + RM_{host}}$ and convert it to the host rest frame, $|\mathrm{RM}|_{\rm host} = |\mathrm{RM_{obs}} - \mathrm{RM_{MW}}|\,(1+z)^{2}$, reported in Table~\ref{tab:frb-sample} for the 27 FRBs with published RMs.

\section{Results}\label{sec:results}

\subsection{The $\mathrm{DM_{host}}$--$M_\star$ Relation}

Our sample consists of 28 non-repeating and 9 repeating FRBs, spanning $\log_{10}(M_\star/M_\odot) \approx 8$--$11.4$ and rest-frame
$\mathrm{DM}_{\rm host} \approx 7$--$1300$~pc\,cm$^{-3}$. Three of the repeaters are associated with a persistent radio source (PRS), which we define as a compact persistent source confirmed at milliarcsecond resolution. Reported PRS candidates associated with FRB\,20201124A and FRB\,20240114A lack VLBI confirmation \citep{bruniNebularOriginPersistent2024, bruniDiscoveryPersistentRadio2025} and are not counted.

\begin{figure*}[ht!]
    \centering
    \includegraphics[width=0.9\linewidth]{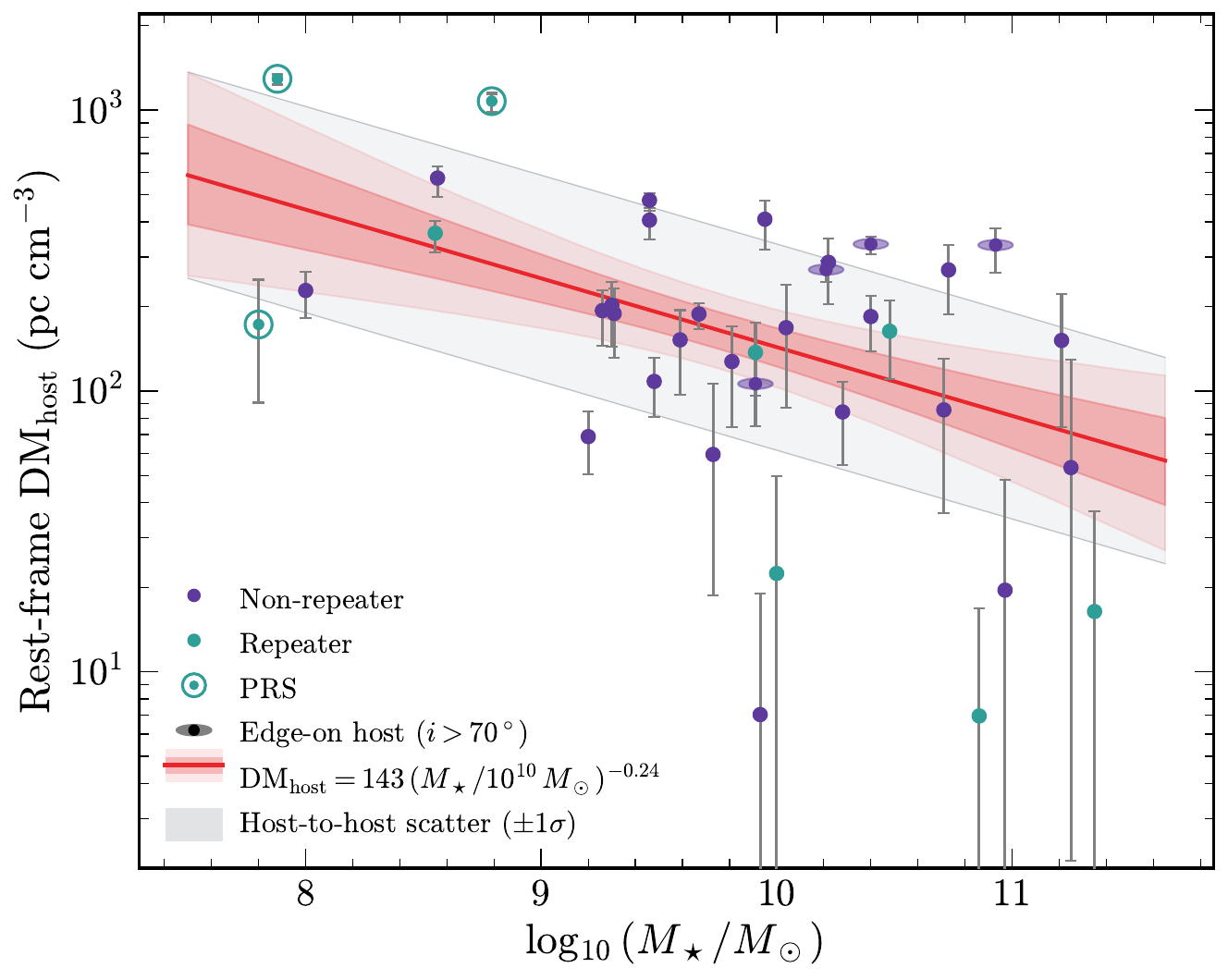}
    \caption{Rest-frame host dispersion measure ($\mathrm{DM_{host}}$) as a
    function of host stellar mass ($M_\star$) for the \sample\ FRBs in our
    sample. The \textit{purple} and \textit{teal} points show non-repeating and repeating FRBs
    respectively. Hosts with an associated PRS are marked by an open circle.
    Hosts observed at high inclination (edge-on, $i > 70^\circ$) are marked by an
    ellipse. The red line shows the best-fit power law, $\mathrm{DM_{host}} = A\,(M_\star/10^{10}\,M_\odot)^{\alpha}$, with
    $A = 143~\mathrm{pc\,cm^{-3}}$ and $\alpha = -0.24$; the dark and light red shaded regions show the corresponding 68\% and 95\% credible
    intervals derived from the posterior samples of $(A,\alpha)$ (Appendix~\ref{sec:appendix}). The gray shaded band shows the intrinsic
    host-to-host scatter about the best-fit relation, $\sigma_{\rm int} = 0.37$~dex.} 
    \label{fig:dmhost}
\end{figure*}

In Figure~\ref{fig:dmhost} we show the rest-frame host dispersion measure ($\mathrm{DM}_{\rm host}$) as a function of host stellar mass. We find a statistically significant anti-correlation between these quantities (Pearson $r = -0.5$, $\log_{10}p = -2.8$; Table~\ref{tab:stats}). The six FRBs with a negligible host DM (Table~\ref{tab:frb-sample}) are plotted at the medians of their $\mathcal{P}(\mathrm{DM_{host}})$ posteriors, with the lower error bar extending to the 2.5th percentile of the posterior. The host-galaxy contribution to the dispersion measure decreases systematically with increasing stellar mass, suggesting underlying physical differences in the local/galactic environments of FRBs.

\begin{figure*}
    \centering
    \includegraphics[width=0.48\linewidth]{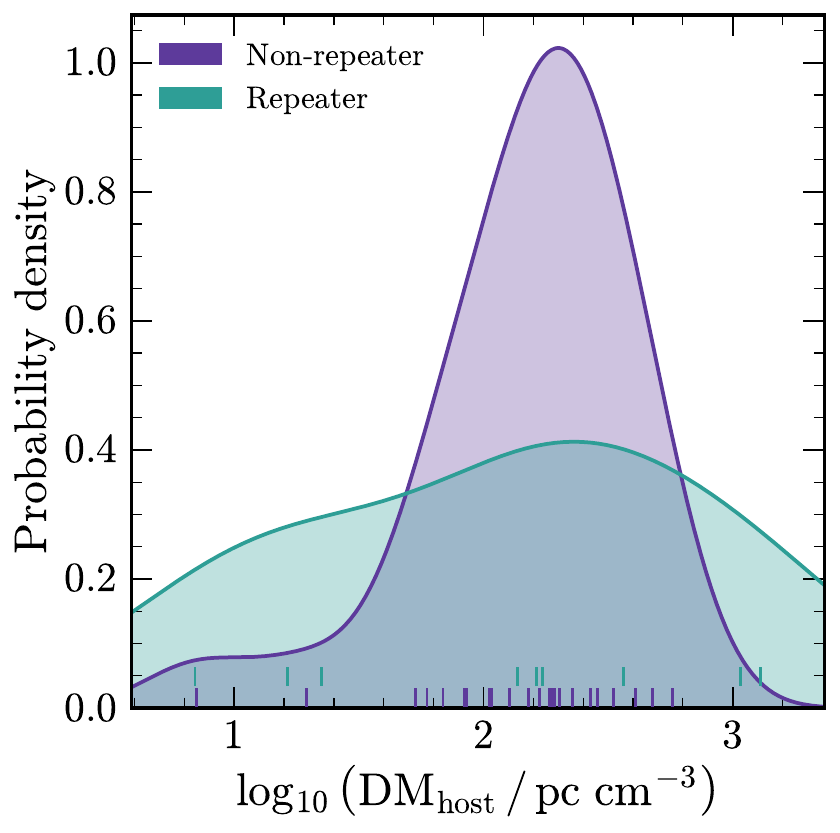}
    \hfill
    \includegraphics[width=0.48\linewidth]{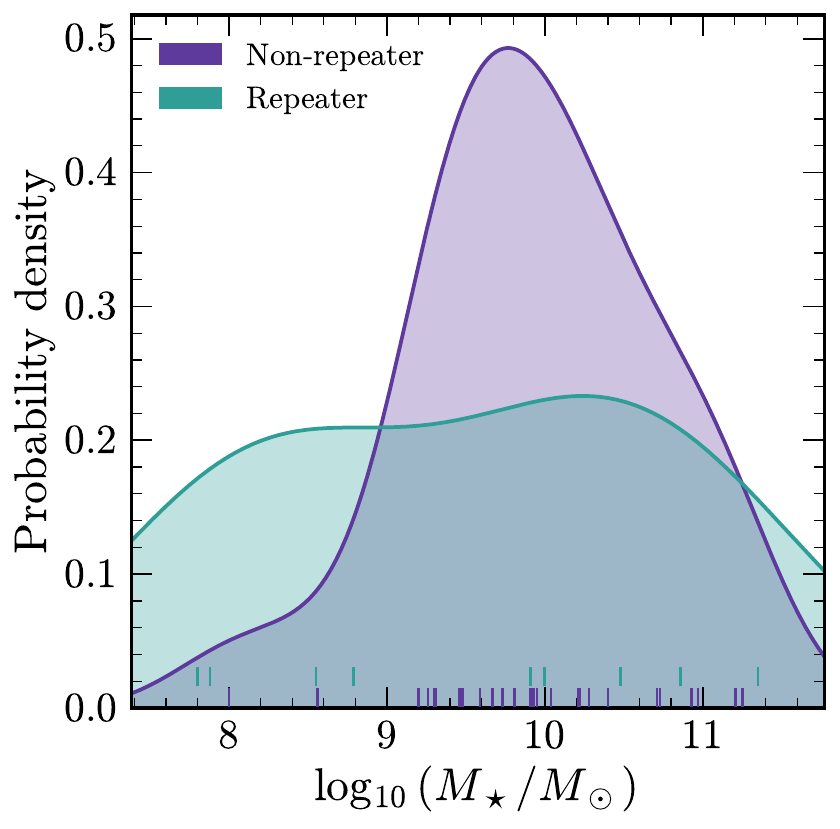}
    \caption{Kernel density estimates for the 28 non-repeating \textit{(purple)} and
    9 repeating \textit{(teal)} FRB hosts shown in Figure~\ref{fig:dmhost}, with
    tick marks along the bottom of each panel showing the individual values. 
    \textit{(Left)} The distribution of rest-frame
    $\mathrm{DM_{host}}$, evaluated in $\log_{10}(\mathrm{DM_{host}})$ with
    Scott's-rule bandwidths of 0.21~dex (non-repeaters) and 0.51~dex
    (repeaters). Each source is represented by the median of its marginalized $P(\mathrm{DM_{host}})$ posterior. \textit{(Right)} The distribution of host stellar mass,
    $\log_{10}(M_\star/M_\odot)$, with Scott's-rule bandwidths of 0.39~dex
    and 0.84~dex, respectively.}
    \label{fig:dmhost-mstar-dist}
\end{figure*}

We find that the anti-correlation is driven by the low-mass and high-mass end of the sample (Table~\ref{tab:stats}). Several of the lowest host galaxy DMs (marked in Table~\ref{tab:frb-sample}) are from $M_\star \gtrsim10^{10}\mathrm{M}_\odot$ hosts. Massive galaxies with large $\rm DM_{host}$ are often edge-on galaxies or scattered, as expected. The largest $\mathrm{DM}_{\rm host}$ sources reside in dwarf galaxies, and repeating FRBs in particular are over-represented among the lowest-mass hosts (Figure~\ref{fig:dmhost-mstar-dist}). If only mid-mass galaxies ($9.0 \leq \log_{10}(M_\star/M_\odot) \leq 10.5$) are included, the correlation is absent ($r = -0.03$), whereas removing edge-on hosts strengthens it ($r = -0.57$; Table~\ref{tab:stats}).

The apparent contradiction with \citet{leung_stellar_2025} arises because we use all non-cluster low-$z$ FRBs. They sought to isolate the CGM contribution and therefore excluded highly scattered FRBs and edge-on hosts. We seek to understand total host DM, including $\rm DM_{loc}$, and do not exclude such galaxies, which tend to be more highly dispersed (see \S\ref{sec:discussion}). When separated into repeating and non-repeating FRBs, the repeaters extend to both higher and lower $\rm DM_{host}$, whereas non-repeaters peak near $\sim150$\,pc\,cm$^{-3}$ (Figure~\ref{fig:dmhost-mstar-dist}). However, the sample is currently too small to draw causal conclusions based on repetition alone.

\subsection{The Power-Law Model} \label{sec:modeling}

To measure the relation between the rest-frame $\mathrm{DM}_{\mathrm{host}}$ and host stellar mass,
we fit a power-law to the $\mathrm{DM}_{\rm host}$ and stellar mass values in the following way. We assume $\mathrm{DM}_{\rm host} = A\,(M_\star/M_{\rm ref})^{\alpha}$, where $M_{\rm ref} = 10^{10}\,M_\odot$ is a reference stellar mass, and we allow for a lognormal intrinsic host-to-host scatter $\sigma_{\rm int}$ (in dex) about the relation. We carry out the fit in log space, where $\log_{10}\mathrm{DM}_{\rm host}$ is normally distributed about $\log_{10}A + \alpha \log_{10}(M_\star/M_{\rm ref})$ with standard deviation $\sigma_{\rm int}$. For each FRB, we transform its measured $\mathcal{P}(\mathrm{DM}_{\rm host})$ into a distribution in $\log_{10}\mathrm{DM}_{\rm host}$ and convolve it with the intrinsic-scatter kernel, so that the per-FRB likelihood is

\begin{equation} \label{eq:like-i}
    \mathcal{L}_i = \int \mathcal{P}(\ell)\,
    \mathcal{N}\!\left(\ell \mid \mu_i,\ \sigma_{\rm int}\right) d\ell ,
\end{equation}
\noindent where $\ell \equiv \log_{10}\mathrm{DM_{host}}$ and
$\mu_i = \log_{10}A + \alpha\,\log_{10}(M_{\star,i}/M_{\rm ref})$.

\noindent The likelihood for the sample is then,

\begin{equation}
    \mathcal{L} = \prod_i^{N_{\rm FRB}} \mathcal{L}_i,
\end{equation}

\noindent which reduces to evaluating each posterior at the model prediction in the limit $\sigma_{\rm int} \rightarrow 0$. We adopt weak uniform priors on $\log A$, $\alpha$, and $\sigma_{\rm int}$, and sample the posterior over $(\log A, \alpha, \sigma_{\rm int})$ using the affine-invariant Markov chain
Monte Carlo ensemble sampler \texttt{emcee} \citep{emcee} with $32$ walkers. We report the median of each marginalized posterior as the best-fit value and the $16$th--$84$th percentiles as the $68\%$ credible intervals. 

The marginalized posteriors give $A = 143^{+24}_{-21}~\mathrm{pc\,cm^{-3}}$, $\alpha = -0.245^{+0.072}_{-0.075}$, and $\sigma_{\rm int} = 0.367^{+0.061}_{-0.049}$~dex, with $P(\alpha < 0) = 0.999$. The posterior distributions are shown in Figure~\ref{fig:corner} (Appendix~\ref{sec:appendix}). The fitted relation implies that the median $\mathrm{DM}_{\rm host}$ declines from $\approx$500\,pc\,cm$^{-3}$ at $\log_{10}(M_\star/M_\odot) = 7.8$ to
$\approx$65\,pc\,cm$^{-3}$ at $\log_{10}(M_\star/M_\odot) = 11.4$, a factor of $7.6^{+6.6}_{-3.4}$ across the mass range of our sample, with a host-to-host scatter of a factor of $\sim$2.3 ($1\sigma$) about the relation at fixed mass.

\input{corr-stats}


\subsection{$|\mathrm{RM}|_{\rm host}$ and Stellar Mass} \label{sec:rm}

RM is the path integral of $n_e B_\parallel$, which can be decomposed in a manner similar to DM. Unlike dispersion, Faraday rotation of extragalactic sources tends to be dominated by the host galaxy and the Milky Way. The diffuse CGM and IGM have magnetic field strengths that are too low to be detectable without stacking large samples \citep{Lan_2020}. 

\begin{figure*}[ht!]
    \centering
    \includegraphics[width=0.9\linewidth]{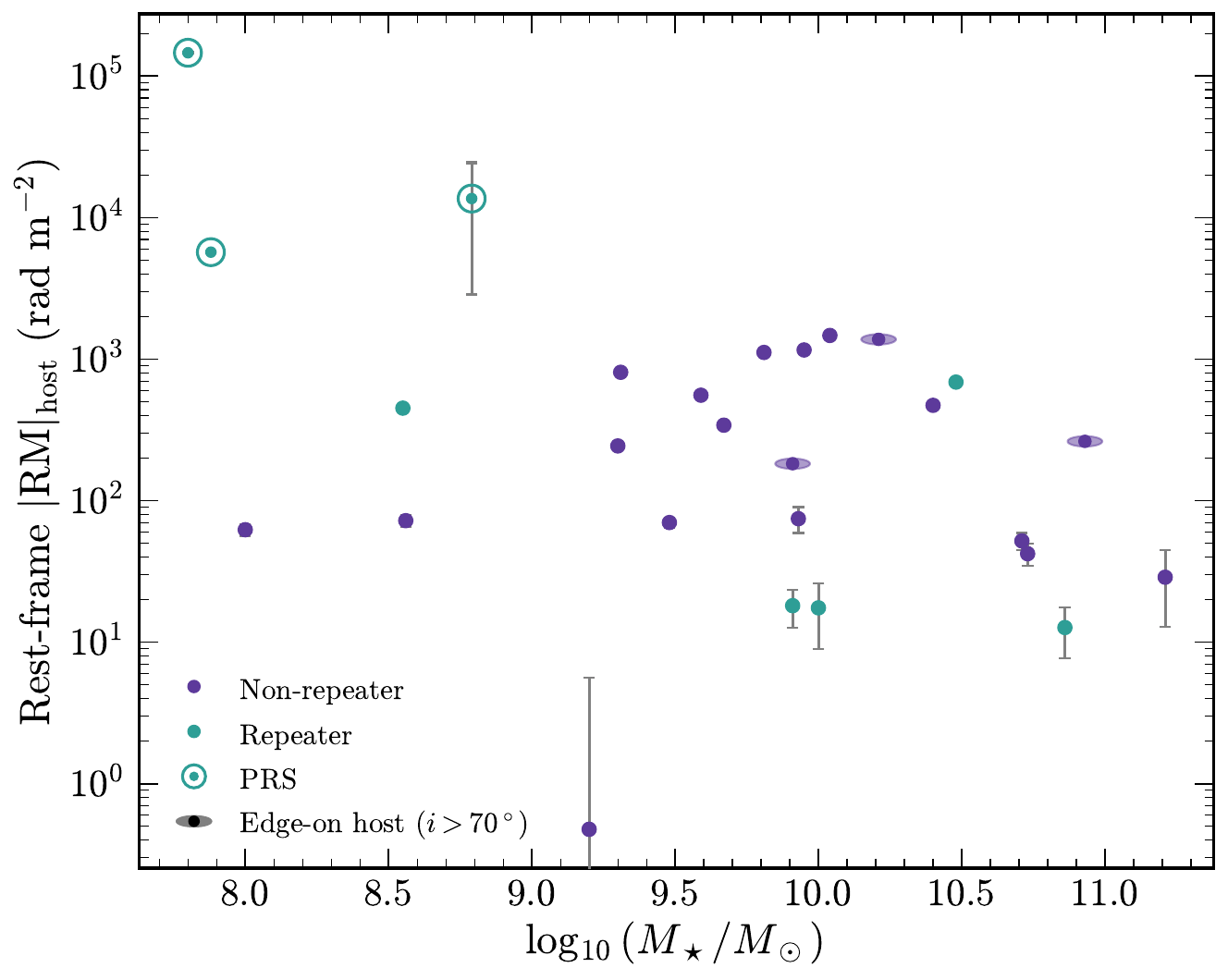}
    \caption{Rest-frame absolute host rotation measure ($|\mathrm{RM}|_{\rm host}$), as a function of host stellar mass ($M_\star$) for the 27 FRBs in our sample with an RM measurement. The \textit{purple} and \textit{teal} points show non-repeating and repeating FRBs
    respectively. Hosts with an associated PRS are marked by an open circle.
    Hosts observed at high inclination (edge-on, $i > 70^\circ$) are marked by an
    ellipse.}
    \label{fig:rm}
\end{figure*}

In Figure~\ref{fig:rm} we plot the absolute value of rest-frame host RM against stellar mass for the 27 FRBs with the required measurements
(\S\ref{sec:data-analysis}). We find a much broader range of $|\mathrm{RM}|_{\rm host}$ values (more than four orders of magnitude, vs.\ $\sim$2 for $\mathrm{DM}_{\rm host}$) over the same stellar-mass range, due to the wide distribution in ordered magnetic field strengths. The dominance of dwarf host galaxies is also more extreme in the case of $\lvert{\rm RM}\rvert$. A Pearson test gives $r = -0.4$ ($\log_{10}p = -1.4$) for the full RM sample, but the
correlation vanishes when the three PRS-hosting FRBs are excluded ($r = -0.01$; Table~\ref{tab:stats}). 

Interestingly, the dwarf host of the non-repeating FRB\,20230708A is an example of a low-mass galaxy with modest RM \citep{muller26}. This source does not have a PRS, despite deep continuum observations ($L_{7.5\,\rm GHz} <\rm 5.2 \times 10^{27} \, erg\,s^{-1} Hz^{-1}$ at 3$\sigma$) \citep{muller26}. Its host is $\sim10\times$ lower in specific star formation than the dwarf hosts of FRB\,20121102A and FRB\,20190520B, which are both repeaters that have high local RMs and PRSs. 

\section{Discussion}\label{sec:discussion}

As shown in \S\ref{sec:results}, the host dispersion measure--stellar mass relation (\dmm) extends across the full mass range of our sample, $\log_{10}(M_\star/M_\odot) \approx 8$--$11.4$. If host galaxies are not masked based on inclination angle and scattering (i.e. edge-on galaxies and highly scattered FRBs are included in the sample), then most of the statistical power (Pearson $r = -0.5$, $\log_{10}p = -2.8$) in the trend arises from the low-mass and high-mass ends of the distribution. Dwarf hosts are overrepresented in the high-$\rm DM_{host}$ and high-$\rm \lvert RM_{host}\rvert$ FRBs; massive galaxies, including the elliptical host of FRB\,20240209A have some of the lowest host DM values, despite being embedded in more massive halos.

We find that the \dmm\ relation does not arise from host galaxy CGM, nor from the host galaxy ISM along typical sightlines. We conclude that the trend is most plausibly set by the immediate environment of the source. Below we step through explanations for these correlations, including galaxy morphology, scattering-related selection effects, the CGM, and host-mass trends in preferred FRB environments. 

\subsection{Ruling out CGM}
We find that the \dmm\ relation does not arise from the host CGM. \citet{leung_stellar_2025} first reported this anti-correlation and interpreted it as a probe of the CGM and of baryonic feedback, and for that reason restricted their sample to face-on $L^*$ star-forming galaxies, excluding dwarfs, ellipticals, edge-on disks, and cluster members.

When reintroducing dwarf galaxies, the CGM cannot produce the observed low-mass host DMs. Nor is the CGM expected to make up the majority of $\rm DM_{host}$ for Milky Way-like galaxies. FRB constraints on diffuse halo gas show that galaxy halos are gas-poor: the nearby non-repeating FRB\,20220319D limits the Milky Way CGM to $\lesssim 47$\,pc\,cm$^{-3}$ and the total CGM mass to $<10^{11}$\,M$_\odot$ \citep{ravi_deep_2025}, massive nearby hosts appear devoid of halo gas \citep{leung_stellar_2025, mccarty2026}, and analyses of the $\rm DM_{cos}/$redshift relation suggest most of the baryons reside outside of halos \citep{connor_gas-rich_2025}. A dwarf halo, with its shallower potential and smaller gas column, contributes very little $\rm DM_{CGM}$. Even with zero feedback, the CGM is more than an order of magnitude short of the $\gtrsim 10^2$–$10^3$\,pc\,cm$^{-3}$ host DMs measured for $\log M_\star \lesssim 9.5$. For example, the host DMs of FRB\,20190417A or FRB\,20190520B would indicate $\log_{10} M_{\rm CGM} = 10.8$--$11.8$, higher than the maximum expected baryon budget of those galaxies by 0.6–1.6\,dex. For fixed feedback strength, $M_{\rm CGM}$ grows with halo mass and therefore with stellar mass, so any halo contribution to $\mathrm{DM}_{\mathrm{host}}$ should decrease toward low $M_\star$, opposite to what we observe. A pure CGM interpretation of the \dmm\ trend requires a strong gradient in feedback vs. stellar mass. A reservoir that grows with $M_\star$ cannot by itself produce an anti-correlation that rises toward low $M_\star$. The same data that correctly constrain the CGM and feedback of $L^*$ galaxies \citep{leung_stellar_2025} therefore cannot explain the high host DMs of the low-mass galaxies in our sample.

\subsection{Ruling out galaxy-averaged ISM}
\subsubsection{The ISM DM}

The disk ISM is also disfavored as the galaxy-averaged property that causes DM$_{\rm host}$ to correlate with $M_\star$. Low-mass galaxies contain less total ionized gas than $L^*$ disks, so a mean ISM column that tracked total gas mass would again decline toward low $M_\star$, in the absence of selection effects. The extreme hosts make this clear. FRB\,20190520B sits in a dwarf of $M_\star \approx 6\times10^8$\,M$_\odot$ yet shows a host DM approaching $\sim 900$\,pc\,cm$^{-3}$ with a dense, structured ionized-gas component \citep{niu_repeating_2022, chen_host_2025}, and FRB\,20190417A is similarly anomalous \citep{moroianu_milliarcsecond_2025}.

\subsubsection{The impact of morphology and geometry}

For a fixed amount of ionized gas in the ISM, DM can vary significantly depending on the galaxy size. Dwarf galaxies have less total ionized ISM than disk galaxies, but in a more compact region. 
Dwarfs are compact and increasingly spheroidal below $M_\star \sim 10^9$\,M$_\odot$ \citep{benavides_disks_2025}, so a sightline near the center of a dwarf threads a short but high-density column with little sightline-to-sightline variance for a uniform distribution of FRB location. For a disk galaxy, most sightlines are out of the galactic plane and therefore receive moderate ISM contribution to the total DM. In the Milky Way, for example, the median sightline has a DM of just 55\,pc\,cm$^{-3}$ from the vantage point of our sun, based on NE2025 \citep{ne2025}, whereas low galactic latitudes can have DMs that exceed $10^3$\,pc\,cm$^{-3}$ \citep{gcmag}. We therefore 
want to test if morphology alone can explain the observed
DM$_{\rm host}$ dependence on $M_\star$.

As a toy example, suppose the ionized ISM mass is $M_{\rm dw, HII}$ and $M_{\rm sp, HII}$ in a dwarf and spiral galaxy, respectively. Assume that the dwarf is a sphere with radius $R_{\rm dw}$ and the spiral is a disk with radius $R_{\rm sp}$ and height $h_{\rm sp}$. The DM from near the center of the dwarf will be $\left < n_e \right >R_{\rm dw}$, so $\mathrm{DM}_{dw} = \frac{M_{\rm dw, HII}}{4/3\pi R_{\rm dw}^3\,\mu_e\,m_p}R_{\rm dw}$ for all sightlines. Out of the plane, the spiral will have $\mathrm{DM}_{sp} = \frac{M_{\rm sp, HII}}{\pi R_{\rm sp}^2\,\mu_e\,m_p}$. Here, $\mu_e$ and $m_p$ are the mean number of free electrons per proton and the proton mass, respectively. The ratio of host galaxy dwarf DM to spiral host DM perpendicular to the plane will be,

\begin{equation}
    \frac{\mathrm{DM_{dw}}}{\mathrm{DM_{sp}}} \approx \frac{3\,M_{\rm dw,HII}}{4\,M_{\rm sp, HII}}\left ( \frac{R_{\rm sp}}{R_{\rm dw}} \right )^2,
\end{equation}

\noindent In words, if the disk to dwarf radius squared is much larger than the ratio of gas masses, the dwarf host DM will be larger due to the difference in morphology for high-latitude sightlines. To put numbers to this toy model, consider a dwarf galaxy with stellar mass $10^{8.5}$\,$M_\odot$ and $R_{dw}=1$\,kpc and a Milky Way-like galaxy with $M_\star\approx 10^{11}$\,$M_\odot$ and $R_{sp}=15$\,kpc and $h_{sp}=1$\,kpc. From the star forming main sequence, the specific star formation rate (SSFR) could be a few times larger in the dwarf \citep{Speagle_2014}, meaning $\frac{M_{\rm dw,HII}}{M_{\rm sp, HII}}\approx 0.01$. In this case, the dwarf host DM would be roughly 2 times larger than the ISM contribution from the spiral galaxy at high latitude, despite the latter having much more ionized gas in total. At low latitudes, the spiral galaxy would contribute $\sim9\times$ more DM than the dwarf for $h_{sp}\approx1$\,kpc. This is an illustrative but contrived example in which the dwarf and spiral gas distributions are presumed to be uniform and the FRB locations are taken to be near the center of their galaxies. The extent to which galaxy morphology impacts the relationship between host/local DM and stellar mass remains unclear. For galaxies on the star-forming main sequence, host DM does not increase significantly with total H\,\textsc{ii} mass because of the inverse relationship between stellar mass and specific star formation and the DM boost from smaller ISM volumes. While future modeling efforts should account for this effect, we discount geometry as the leading explanation because it is difficult for the diffuse ionized ISM of star-forming dwarf galaxies to produce $\sim$\,10$^3$\,pc\,cm$^{-3}$, large RMs, and time-variability in both quantities. Unusual source locations are apparently required.

\subsection{Ruling out observational selection effects}

Geometry alone requires a selection effect in the high-DM tail of spiral galaxies, where observers do not detect FRBs that travel directly through their host galactic plane. There is tentative evidence for this in disk galaxies hosts \citep{bardwajorientation}. However, such an explanation requires mass-dependence of the selection effect. We consider the mechanism behind such a detection bias.

High DM FRBs are more difficult to detect in real-time radio surveys due to intra-channel dispersion smearing, large dispersion delays \citep{connor2019}, and the positive correlation between host galaxy DM and scattering. It is unclear why high-DM low-$M_\star$ FRBs would be easier to detect than high-DM FRBs in massive host galaxies. If the scattering/DM relation is different between dwarf and spiral galaxies, this could lead to a DM selection effect that depends on stellar mass. For example, Galactic pulsars at low latitudes exhibit a positive correlation between $\tau$ and DM that rises more steeply than a power-law, with $\log_{10}\tau_{1\,\rm GHz} \approx -6.46+0.154\log_{10}(\mathrm{DM})+1.07(\log_{10}\mathrm{DM})^2$ \citep{Bhat2004}. If the relation in dwarf galaxies were flatter, or purely power-law, then highly dispersed events would be less scattered for a fixed DM, making them easier to detect. Combined with previous arguments from morphology, disk galaxies could provide a wide range of host DMs where sources with large $\rm DM_{ISM}$ in the host are missed, but most $\rm DM_{\rm host}$ in dwarfs are high but detectable. We are skeptical of this explanation for the trends in Figure~\ref{fig:dmhost} and Figure~\ref{fig:rm} for several reasons. 

First, there is no evidence that dwarf galaxies have less scattering per DM. \citet{Faber24} showed that most FRBs, including those with dwarf hosts, fall on the Milky Way's $\tau/\rm DM$ relation, despite a variety of host galaxy types. Second, a scattering selection effect impacts only the upper-right triangle of the \dmm\ space, unless strong morphology arguments are invoked. There is no obvious selection effect against low-DM$_{\rm host}$ dwarf galaxies otherwise. The compact ISM of dwarfs is itself unlikely to produce $\rm DM_{host}\gtrsim500$\,pc\,cm$^{-3}$. Further characterization of the relationship between host galaxy orientation, DM, and scattering will be enabled by upcoming localized FRB samples \citep{hallinan2019dsa2000radiosurvey}.

\subsection{A local circumburst origin}

We find the most likely explanation for these trends is regions of ionized gas in the vicinity of FRBs and the spatial statistics of star formation as a function of galaxy type. We favor scenarios in which low-mass host galaxies preferentially produce FRBs in unusually dense locations. FRBs in massive galaxies appear more likely to reside in globular clusters, at higher offsets, and in old stellar populations far from their birth nebula. The spatial statistics of H\,\textsc{ii} (i.e. clumpy vs. uniform) will also impact the DM of FRBs whose progenitors trace star formation.

\subsubsection{Evidence for the local, circumburst origin}

Extreme cases such as FRB\,20190520B and FRB\,20190417A are difficult to explain without circumburst material very near to the source. FRB\,20121102A requires a compact magnetoionic plasma to provide its RM$\sim$\,10$^5$\,rad\,m$^{-2}$ \citep{michilli_extreme_2018}. Three of the six lowest-$M_\star$ hosts in our sample are associated with a persistent radio source (PRS), and two of these are the highest host DMs known. The three PRS repeaters have the highest FRB RMs measured to date. A PRS traces a dense, magnetized, parsec scale nebula around the engine, as in FRB\,20121102A \citep{tendulkar_host_2017, michilli_extreme_2018} and FRB\,20190520B \citep{niu_repeating_2022, chen_host_2025}, both in low-mass, low-metallicity hosts. In the case  of FRB\,20190520B, \citet{Ocker2023} showed that the host scattering of that source can vary on minutes timescales. The source also shows large swings in RM, with sign flips on timescales of weeks \citep{anna-thomasMagneticFieldReversal2023}. More recently, multi-year observing campaigns of repeaters have shown steadily declining  DMs \citep{Niu2026, pandhi_steadily_2026}. This is suggestive of an expanding supernova remnant \citep{connor2016, piro2018} or magnetar wind nebula, though the origin and future trends of DM$(t)$ remains an open question. Finally, long-term monitoring of the FRB\,20121102A PRS probes the same evolving local medium \citep{bhardwaj_constraining_2025}. The CGM and the galaxy-averaged ISM cannot vary on these timescales; only local material in which the source is embedded can be this dynamic.

The $\lvert$RM$\rvert$–Stellar Mass data point to the same conclusion. An appreciable $\lvert$RM$\rvert$ requires ionized gas and a coherent magnetic field threading it, which is met in a dense circumburst nebula but not in a galaxy halo. That host $\lvert$RM$\rvert$ is highest in the lowest $M_\star$, in step with $\mathrm{DM}_{\mathrm{host}}$, therefore reinforces a local origin, with the extreme magneto-ionic environments of FRB\,20121102A and FRB\,20190520B \citep{michilli_extreme_2018, chen_host_2025} and the variable RM of FRB\,20220529A \citep{li_sudden_2026} as examples. 

\subsubsection{Star-formation statistics}

The existence of such high density regions with FRBs in low-mass hosts may be related to the statistics of star formation and the spatial distribution of H\,\textsc{ii} across galaxies of different masses. Beyond PRS FRBs, the \dmm\ anti-correlation extends over several orders of magnitude in stellar mass, including  apparent non-repeaters with normal RM and scattering properties. This could be caused by the statistical properties of star formation as a function of metallicity and SSFR in FRB host galaxies.

In a sample of galaxies with $8 \leq \log_{10}(M_\star/M_\odot) < 10.3$ at $0.064 < z < 0.1$, \citet{mintz2024} found that low-mass galaxies with high SSFR have $H\alpha$ emission that is heterogeneous and compact. Less active galaxies in their sample had diffuse $H\alpha$ emission that was less peaky and more evenly distributed. We note that the SSFR of FRB\,20230708A is lower than other dwarfs in this sample and its host DM is only $\sim$200\,pc\,cm$^{-3}$ and its RM is modest. It does not have a PRS.

Combining these arguments, the gradual trend across the full mass range in our sample may be caused by the fact that SSFR is inversely proportional to stellar mass, leading to a smaller fraction of FRB engines embedded in dense H\,\textsc{ii} regions. Star forming regions in massive hosts have higher metallicities and may be less clumpy, leading to lower DM from the immediate environment. At the highest $M_\star$, FRBs appear to come from older populations whose natal nebula dissipated long before the source was detected. This assumes that $M_\star$ corresponds to the parent galaxy and not the globular cluster or satellite, as in the case of FRB\,20200120E, which resides in a globular cluster \citep{kirsten_repeating_2022}, and FRB\,20240209A, which is offset 40\,kpc from a massive elliptical \citep{eftekhari_massive_2025, shah_repeating_2025}. Both of those sources align with the picture of SSFR and metallicity being ultimately responsible for DM$_{\rm loc}$. Star forming dwarf galaxies, on the other hand, have higher SSFR and clumpier, denser H\,\textsc{ii} regions, with faint extended H$\alpha$ that standard tracers miss \citep{lee_deeper_2016}. 

Long gamma ray bursts (LGRBs) and superluminous supernova (SLSN) are preferentially found in low-metallicity star forming dwarf galaxies, often at UV-bright regions within the host \citep{metzger2017}. While other hosts are not precluded for those sources \citep{Nicholl_2017}, dwarf starbursts are apparently efficient at producing young, low-metallicity, high angular momentum sites of massive stellar death. By analogy, star forming dwarf galaxies that host FRBs may preferentially produce similar magnetars, leaving young repeating FRBs with coincident nebulae.

\subsection{Connection to the deficit of low-mass hosts}

In \citet{sharma_preferential_2024}, the authors use a sample of host galaxy properties to show that FRBs trace star formation (as opposed to, e.g., stellar mass), but with a marked deficit of $M_\star \lesssim 10^{9}\,M_\odot$ hosts compared to global star formation and the host galaxies of core-collapse supernovae. This could emerge if FRBs were a biased tracer of star formation, perhaps caused by a positive metallicity dependence on FRB occurrence rate. Another explanation is that low-mass hosts produce higher-DM$_{\rm host}$ events, leading to a selection bias in FRB surveys (low max search DM, intrachannel dispersion smearing, scattering, etc.).

We have shown that dwarf galaxies are more likely to have large DM$_{\rm host}$. It is unlikely that $\sim$\,10$^3$\,pc\,cm$^{-3}$ is a ceiling for host DM. $\rm DM>3\times10^3$\,pc\,cm$^{-3}$ would exceed the maximum search DM of many surveys, with deleterious effects starting around $10^3$\,pc\,cm$^{-3}$. If this effect is significant and related to the reported deficit of dwarf hosts compared to core-collapse supernovae, it may still be downstream of metallicity bias \citep{sharma_preferential_2024}. But it would emerge at the time of detection rather than in the intrinsic FRB rate per unit star formation across different galaxies.

\subsection{Relevance to FRB cosmology}

Our result has implications for the emerging field of FRB cosmology \citep{frbcos}. When FRBs are used to measure the cosmic baryons \citep{Macquart2020, flimflam, connor_gas-rich_2025, sharma_probing_2025, wang25, Hussaini}, the host DM contribution is a primary source of noise, acting as a nuisance parameter that must be marginalized over or fit out. 

Therefore, if massive host galaxies (e.g. $\log M_\star \gtrsim 9.5$) have relatively little local DM contribution, low-mass hosts can be avoided or down-weighted in analyses that use FRBs to measure cosmic gas. 

For example, upcoming radio surveys such as the DSA \citep{hallinan2019dsa2000radiosurvey} and CHORD \citep{Vanderlinde19} will localize tens of thousands of FRBs, measuring DM, RM, and scattering time. LSST will find billions of galaxies within redshift 1.5 \citep{Ivezic}, some of which will be FRB host galaxies. Photometry would offer an initial estimate of host galaxy stellar mass and those FRBs could provide a pristine subset with which to carry out proposed FRB cosmology experiments. Limited spectroscopic follow-up resources could be devoted to massive hosts identified in photometric data; dwarfs and other low-mass hosts should be avoided for these science cases due to added noise and bias.

\section{Conclusions}\label{sec:conclusions}

We have investigated the relationship between stellar mass and FRB propagation effects in their host galaxies. We study $\rm DM_{host}$ and $|\mathrm{RM}|_{\rm host}$ for FRB sources at $z \lesssim 0.25$ with secure host association, excluding FRBs that reside in galaxy clusters. In the absence of mass-dependent FRB environments or strong selection effects, the natural expectation is that $\rm DM_{host}$ should increase with $M_\star$.

We show an anti-correlation between rest-frame host DM and stellar mass in Figure~\ref{fig:dmhost}. The lowest-mass hosts contain several of the highest host-DM events, including sources associated with persistent radio emission, while the most massive systems tend to have relatively low DM contributions. The host dispersion measure--stellar mass relation (\dmm) is tracing very local FRB environments, or a connection between low-mass galaxies and unusually dense star-forming or circumburst regions. This result points to the emergent need for studying the local sub-parsec scale environments of FRBs to constrain their origins. Our finding is consistent with \citet{leung_stellar_2025}, but our sample spans the full range of galaxy types and inclination angles, leading us to a new interpretation. We also searched for a trend between $|\mathrm{RM}|_{\rm host}$ and host stellar mass. While we do not find evidence for an anti-correlation spanning the full mass range, it is notable that the three largest 
$|\mathrm{RM}|_{\rm host}$ values come from repeaters with PRSs in low-mass galaxies.

Host DM and absolute RM are composed of the CGM, the ISM, and the local environment. The former two are not expected to decrease as a function of host stellar mass. We note that while the typical FRB host DM is likely dominated by the ISM and CGM, negative trends between those quantities and host mass are not plausible across $10^{8}\,M_\odot<M_\star<10^{11.4}\,M_\odot$. Thus, the observed anti-correlation is likely due to the local environment of FRBs, where there is an apparent preference for denser ionized environments at low-$M_\star$. Massive galaxies may have more ionized gas in total, but conditional on forming an FRB, sources appear to occur in lower-density regions, older stellar populations, globular clusters, or cleaner sightlines. If the correlation is intrinsic and not an observational bias, $M_\star$ serves as a proxy for an underlying variable  that influences the progenitor channel and the resulting immediate environment. We consider metallicity to be the most likely culprit, leading to delayed channels in massive hosts and dense, young environments in dwarf galaxies.

The results are as follows:

\begin{itemize}
    \item We find a statistically significant anti-correlation between $\mathrm{DM}_{\rm host}$ and host stellar mass (Pearson $r\approx-0.5$, $\log_{10}p\approx-2.8$; Table~\ref{tab:stats}), such that lower-mass galaxies host FRBs with systematically higher $\mathrm{DM}_{\rm host}$.

    \item The trend spans the full mass range of the sample and is driven statistically by both the low- and high-mass ends. It cannot be explained by the CGM, the galaxy-averaged ISM, or galaxy morphology alone.

    \item We find no evidence for a trend between $|\mathrm{RM}|_{\rm host}$ and host stellar mass across the full mass
    range (Pearson $r\approx-0.4$, $\log_{10}p\approx-1.4$; Table~\ref{tab:stats}).

    \item The trend in \dmm\ is best explained by the local ($\lesssim$\,100\,pc) environment around FRBs. The causal origin of this relation is plausibly the spatial statistics of star formation and a metallicity-dependence on FRB progenitors: high-SSFR dwarf galaxies have clumpy, heterogeneous H\,\textsc{ii}, leading to higher local DMs than Milky Way-like galaxy hosts for sources that are embedded in star forming regions; high-metallicity galaxies (e.g. the massive quiescent host of FRB\,20240209A) are more likely to form FRBs in older stellar populations.
\end{itemize}

Our interpretation makes several testable predictions. First, larger samples of well-localized local Universe FRBs will improve statistics and better characterize the trend. Upcoming surveys such as CHIME/VLBI \citep{chimevlbi}, CHORD \citep{Vanderlinde19}, and DSA \citep{hallinan2019dsa2000radiosurvey} will localize thousands of FRBs at $\lesssim200$\,mas. All-sky monitors such as CASM \citep{connor2026256antennacoherentallskymonitor}, BURSTT \citep{Lin_2022}, and CHARTS \citep{charts} will preferentially find nearby sources whose host environments can be studied in detail. Trends between radio propagation effects (host DM, RM, scattering) and SSFR, metallicity, morphology, and stellar mass can be explored in these upcoming datasets. Our model predicts that FRBs in low-metallicity, low-mass hosts are more likely to be coincident with peaks in the spatial distribution of star formation. It also predicts that high-mass hosts are less likely to have PRSs. With just a few PRSs discovered to date, it is difficult to make statistical claims about their prevalence. Fortunately, the DSA will survey $\sim$\,30,000\,deg$^2$ to sub-$\mu$Jy RMS, offering a blind survey of compact, luminous radio sources \citep{hallinan2019dsa2000radiosurvey}. Next, our skepticism about the
$\rm DM_{host}$ trend being due to a selection effect can be tested. If there exists an observational bias against high-DM events from high-mass galaxies and low-DM events in low-mass galaxies, this can be addressed with $\rm DM_{host}$ / scattering studies and improved real-time FRB detection characterization and completeness. Finally, a local origin to the anti-correlation between $\rm DM_{host}$ and host stellar mass predicts that propagation measures of repeating FRBs with high DM will vary more rapidly than repeaters with low DMs. This includes $\rm RM_{host}$, $\rm DM_{host}$, and scattering times.

Optical and infrared followup will also play a central role in understanding the immediate host environment. Localization improvements in FRB host galaxies have led to an opportunity to study sub-kiloparsec  environments. The growing sample of localized FRBs enables statistical tests that were not possible only a few years ago. Recent deep JWST imaging of FRB\,20250316A \citep{blanchard_james_2025} has already demonstrated the power of this approach, revealing a resolved stellar environment and a faint near-infrared source near the FRB position. This approach points to a new path forward: high-angular resolution imaging and spectroscopy of the nearest and best-localized sources, allowing for 
direct diagnostics about the local environment of individual FRBs.


\appendix
\section{Power-Law Fit Posterior Distributions} \label{sec:appendix}
\restartappendixnumbering

\begin{figure}[htbp]
\centering
\includegraphics[width=\columnwidth]{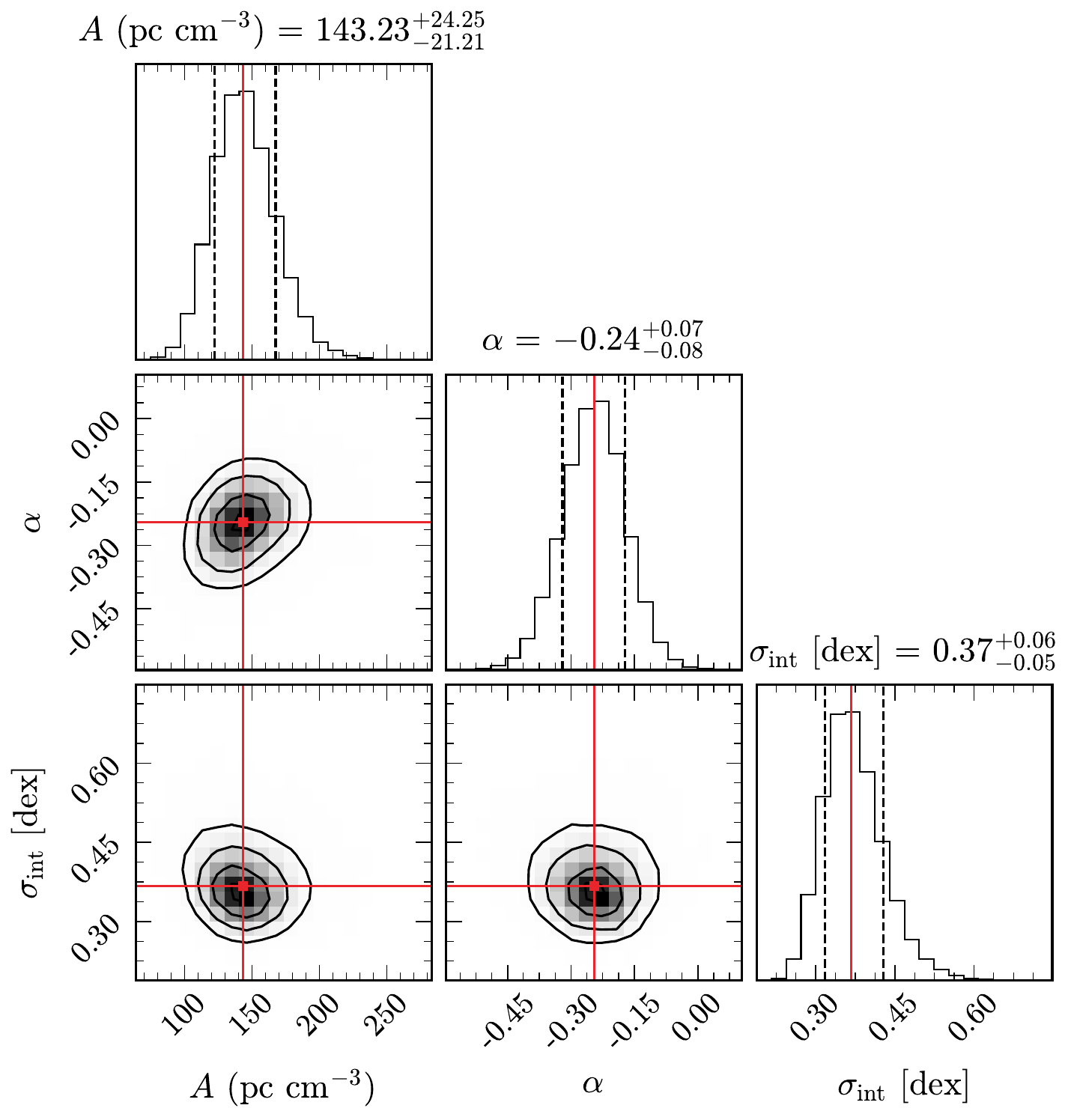}
\caption{Posterior probability distributions for the
$\mathrm{DM_{host}}$--$M_\star$ power-law normalization $A$ (in
pc\,cm$^{-3}$), power-law index $\alpha$, and intrinsic (host-to-host)
log-normal scatter $\sigma_{\rm int}$ (in dex). Contours show the
1$\sigma$ and 2$\sigma$ credible regions of the joint posterior; dashed
lines and titles above each 1D marginal panel mark the median and 68\%
credible interval. The red lines mark the adopted median values,
$A = 143.2~\mathrm{pc\,cm^{-3}}$, $\alpha = -0.245$, and
$\sigma_{\rm int} = 0.367$~dex, used in Figure~\ref{fig:dmhost}.}
\label{fig:corner}
\end{figure}

As described in \S\ref{sec:modeling}, we fit the $\mathrm{DM_{host}}$--$M_\star$ relation with \texttt{emcee} \citep{emcee} using 32 walkers. After a 2000-step burn-in (discarded) and a 5000-step production run, the chains yielded 128{,}000 posterior samples with a mean acceptance fraction of 0.64 and an integrated autocorrelation time of $\tau \approx 36$--$41$ steps, corresponding to $\sim$3100 effective independent draws. Figure~\ref{fig:corner} shows the resulting posterior distributions. The medians and 68\% credible intervals are $A = 143^{+24}_{-21}~\mathrm{pc\,cm^{-3}}$, $\alpha = -0.245^{+0.072}_{-0.075}$, and $\sigma_{\rm int} = 0.367^{+0.061}_{-0.049}$~dex (a fractional intrinsic scatter of $\sim$130\%).

\section*{Acknowledgments}

We thank Danielle Frostig for her comments on the manuscript. We also thank Casey Law and Vadim Semenov for useful discussions on host galaxy DM in dwarfs etc. 
The Berger Time-Domain research group at Harvard is supported by the NSF and NASA. This research has made use of the NASA Astrophysics Data System (ADS), the NASA/IPAC Extragalactic Database (NED), and NASA/IPAC Infrared Science Archive (IRSA, which is funded by NASA and operated by the California Institute of Technology).

\software{Astropy \citep{astropy}, Matplotlib \citep{matplotlib}, NumPy \citep{numpy}, SciPy \citep{scipy}, pandas \citep{pandas}, emcee \citep{emcee}, corner \citep{corner}, PyGEDM \citep{pygedm}}

\bibliography{references}
\bibliographystyle{aasjournalv7}

\end{document}

%% file: frb-sample.tex
\begin{deluxetable*}{c|ccccccccccc}
\tabletypesize{\scriptsize}
\setlength{\tabcolsep}{2pt}
\tablecaption{FRB Sample\label{tab:frb-sample}}
\tablehead{\colhead{FRB} & \colhead{R.A.} & \colhead{Decl.} & \colhead{$z$} & \colhead{$\mathrm{DM}_{\mathrm{obs}}$} & \colhead{$\log_{10}(M_\star/M_\odot)$} & \colhead{$\mathrm{DM}_{\mathrm{host}}$\tablenotemark{$\dagger$}} & \colhead{$\mathrm{RM}_{\mathrm{obs}}$} & \colhead{$|\mathrm{RM}|_{\mathrm{host}}$\tablenotemark{$\ddagger$}} & \colhead{Repeater} & \colhead{PRS} & \colhead{References}\\
\colhead{} & \colhead{(deg)} & \colhead{(deg)} & \colhead{} & \colhead{(pc cm$^{-3}$)} & \colhead{} & \colhead{(pc cm$^{-3}$)} & \colhead{(rad m$^{-2}$)} & \colhead{(rad m$^{-2}$)} & \colhead{} & \colhead{} & \colhead{}}
\startdata
20200430A   & 229.7065    & 12.3763     & 0.161      & 380          & 9.30    & $202^{+43}_{-58}$                     & $195.3\pm 0.7$         & $244 \pm 7$        & No       & No   & 1, 2, 3, 6 \\
20210807D   & 299.2214    & -0.7624     & 0.12927    & 251.3        & 10.97   & $20^{+29}_{-16}$\tablenotemark{$\ast$}& \nodata                & \nodata            & No       & No   & 2, 8 \\
20211127I   & 199.8088    & -18.8379    & 0.0469     & 234.97       & 9.48    & $108^{+23}_{-27}$                     & $-67\pm1$              & $70 \pm 6$         & No       & No   & 2, 6, 8 \\
20211212A   & 157.3508    & 1.3604      & 0.0707     & 209          & 10.28   & $84^{+24}_{-30}$                      & \nodata                & \nodata            & No       & No   & 2, 8 \\
20220725A   & 353.3152    & -35.9903    & 0.1926     & 290.4        & 10.71   & $86^{+45}_{-49}$                      & $-26.3\pm0.7$          & $52 \pm 7$         & No       & No   & 4, 6, 8 \\
20220920A   & 240.2571    & 70.9188     & 0.1582     & 315          & 9.81    & $127^{+42}_{-53}$                     & $-830.3\pm8.3$         & $1115 \pm 13$      & No       & No   & 5, 13, 15  \\
20230526A   & 22.2326     & -52.7173    & 0.157      & 361.4        & 9.31    & $189^{+42}_{-57}$                     & $613\pm2$              & $807 \pm 7$        & No       & No   & 4, 6, 8 \\
20230628A   & 166.7867    & 72.2818     & 0.127      & 345.15       & 9.26    & $193^{+35}_{-48}$                     & \nodata                & \nodata            & No       & No   & 4, 13 \\
20231226A   & 155.3637    & 6.1103      & 0.1569     & 329.9        & 9.59    & $152^{+42}_{-55}$                     & $428.4\pm0.3$          & $556 \pm 7$        & No       & No   & 4, 6, 8 \\
20240210A   & 8.7796      & -28.2708    & 0.023686   & 283.73       & 9.67    & $188^{+17}_{-22}$                     & $-325\pm1$             & $341 \pm 5$        & No       & No   & 4, 6, 8 \\
20201124A   & 77.0146     & 26.0607     & 0.0982     & 411          & 10.48   & $163^{+47}_{-53}$                     & $-614\pm2$             & $688 \pm 6$        & Yes      & No   & 20, 22, 7  \\
20190417A   & 294.7745    & 59.3269     & 0.12817    & 1379         & 7.88    & $1292^{+43}_{-57}$                    & $4509\pm18$            & $5701 \pm 24$      & Yes      & Yes  & 9 \\
20121102A   & 82.9946     & 33.1479     & 0.19273    & 558.1        & 7.80    & $172^{+77}_{-81}$                     & $102700\pm 100$        & $146129 \pm 142$   & Yes      & Yes  & 10, 11, 26, 27, 31\\
20190608B   & 334.0199    & -7.8983     & 0.1177805  & 340.05       & 10.40   & $184^{+34}_{-46}$                     & $353\pm2$              & $472 \pm 7$        & No       & No   & 3, 29, 32, 33 \\
20190520B   & 240.5178    & -11.2881    & 0.241      & 1210.3       & 8.79    & $1075^{+71}_{-98}$                    & $8840\pm7000$          & $13633 \pm 10781$  & Yes      & Yes  & 14, 34, 35, 48\\
20210117A   & 339.9792    & -16.1514    & 0.2145     & 728.95       & 8.56    & $573^{+58}_{-83}$                     & $-45.8\pm0.7$          & $72 \pm 7$         & No       & No   & 2, 6, 8, 16\\
20210410D   & 326.0862    & -79.3182    & 0.1415     & 578.78       & 9.46    & $406^{+44}_{-59}$                     & \nodata                & \nodata            & No       & No   & 17 \\
20220912A   & 347.2704    & 48.7066     & 0.0771     & 219.46       & 10.00   & $22^{+27}_{-17}$\tablenotemark{$\ast$}& $-0.08\pm5.39$         & $17 \pm 9$         & Yes      & No   & 18, 19, 36 \\
20240114A   & 321.9160    & 4.3294      & 0.130287   & 527.723      & 8.55    & $364^{+40}_{-53}$                     & $338.1\pm0.1$          & $451 \pm 6$        & Yes      & No   & 21, 37 \\
20240209A   & 289.9160    & 86.0623     & 0.1384     & 176.518      & 11.35   & $16^{+21}_{-13}$\tablenotemark{$\ast$}& \nodata                & \nodata            & Yes      & No   & 23, 24 \\
20210603A   & 10.2741     & 21.2263     & 0.1772     & 500.147      & 10.93   & $331^{+47}_{-67}$                     & $-219.00\pm0.01$       & $262 \pm 7$        & No       & No   & 25 \\
20220207C   & 310.1995    & 72.8823     & 0.0433     & 262.3        & 9.91    & $106^{+28}_{-31}$                     & $162.48\pm0.04$        & $182 \pm 5$        & No       & No   & 5, 13, 15 \\
20240201A   & 149.9056    & 14.0880     & 0.042729   & 374.5        & 10.21   & $270^{+21}_{-27}$                     & $1275.0\pm0.4$         & $1380 \pm 5$       & No       & No   & 6, 8, 38 \\
20231120A   & 143.9840    & 73.2847     & 0.0368     & 438.9        & 10.40   & $333^{+21}_{-26}$                     & \nodata                & \nodata            & No       & No   & 13, 39\\
20250316A   & 182.4347    & 58.8491     & 0.0067     & 161.82       & 9.20    & $69^{+16}_{-18}$                      & $16.79\pm0.85$         & $0 \pm 5$          & No       & No   & 28, 50\\
20220319D   & 32.1779     & 71.0353     & 0.0112     & 110.95       & 9.93    & $7^{+12}_{-6}$\tablenotemark{$\ast$}  & $59.9\pm14.3$          & $75 \pm 15$        & No       & No   & 5, 15, 41\\
20180916B   & 29.5031     & 65.7168     & 0.0337     & 348.76       & 9.91    & $137^{+38}_{-41}$                     & $-114.6\pm0.6$         & $18 \pm 5$         & Yes      & No   & 2, 30, 49 \\
20190714A   & 183.9797    & -13.021     & 0.2365     & 504.13       & 10.22   & $287^{+61}_{-83}$                     & \nodata                & \nodata            & No       & No   & 2, 3, 12 \\
20191001A   & 323.3517    & -54.7483    & 0.234      & 507          & 10.73   & $270^{+62}_{-83}$                     & $51.1\pm0.9$           & $42 \pm 8$         & No       & No   & 2, 3, 6, 40 \\
20220307B   & 350.8745    & 72.1924     & 0.2481     & 499.15       & 10.04   & $168^{+71}_{-81}$                     & $-947.2\pm12.3$        & $1470 \pm 21$      & No       & No   & 5, 13, 15 \\
20220825A   & 311.9815    & 72.5850     & 0.2414     & 651.2        & 9.95    & $409^{+67}_{-91}$                     & $750.2\pm6.7$          & $1161 \pm 13$      & No       & No   & 5, 13, 15  \\
20230124A   & 231.9170    & 70.9681     & 0.0939     & 590.60       & 9.46    & $477^{+30}_{-41}$                     & \nodata                & \nodata            & No       & No   & 13, 39 \\
20221101B   & 342.2158    & 70.6815     & 0.2395     & 490.7        & 11.21   & $151^{+70}_{-77}$                     & $-32.2\pm9.1$          & $29 \pm 16$        & No       & No   & 13, 15, 39 \\
20191228A   & 344.4305    & -29.5941    & 0.243      & 298          & 9.73    & $59^{+47}_{-41}$                      & \nodata                & \nodata            & No       & No   & 1, 3, 39\\
20230708A   & 303.1155    & -55.3563    & 0.105      & 411.51       & 8.00    & $228^{+38}_{-46}$                     & $-7.5\pm0.4$           & $62 \pm 6$         & No       & No   & 6, 8, 42 \\
20200120E   & 149.4779    & 68.8169     & 0.0008     & 87.818       & 10.86   & $7^{+10}_{-6}$\tablenotemark{$\ast$}  & $-29.8\pm0.5$          & $13 \pm 5$         & Yes      & No   & 43, 44, 45, 47 \\
20210405I   & 255.3396    & -49.5452    & 0.066      & 565.17       & 11.25   & $53^{+75}_{-43}$\tablenotemark{$\ast$}& \nodata                & \nodata            & No       & No   & 46 \\
\enddata
\tablenotetext{\ast}{FRBs with host DM consistent with zero.}
\tablenotetext{\dagger}{Calculated in this work (\S\ref{sec:data-analysis}).}
\tablenotetext{\ddagger}{Calculated in this work (\S\ref{sec:data-analysis}).}
\tablecomments{The R.A. and Decl. values are for the FRB and are given in the J2000 system. The calculated DM$_{\mathrm{host}}$ and $|\mathrm{RM}|_{\mathrm{host}}$ are in host rest frame. `Repeater' and `PRS' flag known repeating sources and hosts with an associated persistent radio source.}
\tablerefs{(1)~\citet{bhandariCharacterizingFastRadio2022}; (2)~\citet{gordon_demographics_2023}; (3)~\citet{dayAstrometricAccuracySnapshot2021}; (4)~\citet{leung_stellar_2025}; (5)~\citet{law_deep_2024}; (6)~\citet{scottHightimeresolutionProperties352025}; (7)~\citet{ravi_host_2022}; (8)~\citet{shannonCommensalRealtimeASKAP2025}; (9)~\citet{moroianu_milliarcsecond_2025}; (10)~\citet{bhardwaj_constraining_2025}; (11)~\citet{michilli_extreme_2018}; (12)~\citet{heintz_host_2020}; (13)~\citet{sharma_preferential_2024}; (14)~\citet{chen_host_2025}; (15)~\citet{shermanDeepSynopticArray2024}; (16)~\citet{bhandari_non-repeating_2023}; (17)~\citet{caleb_sub-arcsec_2023}; (18)~\citet{hewitt_milliarcsecond_2024}; (19)~\citet{ravi_deep_2023}; (20)~\citet{kumarCircularlyPolarizedRadio2022}; (21)~\citet{bhardwaj_hyperactive_2025}; (22)~\citet{nimmoMilliarcsecondLocalisationRepeating2022}; (23)~\citet{shah_repeating_2025}; (24)~\citet{eftekhari_massive_2025}; (25)~\citet{cassanelliFastRadioBurst2024}; (26)~\citet{tendulkar_host_2017}; (27)~\citet{marcote_repeating_2017}; (28)~\citet{collaborationFRB20250316ABrilliant2025}; (29)~\citet{dayHighTimeResolution2020}; (30)~\citet{the_chimefrb_collaboration_chimefrb_2019}; (31)~\citet{chatterjee_direct_2017}; (32)~\citet{chittidi_dissecting_2021}; (33)~\citet{bhandariHostGalaxiesProgenitors2020}; (34)~\citet{niu_repeating_2022}; (35)~\citet{anna-thomasMagneticFieldReversal2023}; (36)~\citet{zhangFASTObservationsFRB2023}; (37)~\citet{tianDetectionLocalisationHighly2024}; (38)~\cite{glowackiInvestigationCorrelationsFRB2025}; (39)~\citet{sangUnveilingDistributionRedshift2025}; (40)~\citet{bhandariLimitsPrecursorAfterglow2020}; (41)~\citet{ravi_deep_2025}; (42)~\citet{muller26}; (43)~\citet{bhardwajNearbyRepeatingFast2021}; (44)~\citet{nimmoBurstTimescalesLuminosities2021}; (45)~\citet{deblokHIGHRESOLUTIONROTATIONCURVES2008}; (46)~\citet{driessenFRB20210405INearby2023}; (47)~\citet{kirsten_repeating_2022}; (48)~\citet{Ocker2023}; (49)~\citet{marcote_repeating_2020}; (50)~\citet{kaurHighresolutionGiantMetrewave2026}}
\end{deluxetable*}

%% file: corr-stats.tex
\begin{deluxetable*}{clcCC}
\tabletypesize{\footnotesize}
\tablecaption{Pearson correlation tests of $\mathrm{DM_{host}}$ and
$|\mathrm{RM}|_{\rm host}$ against host stellar mass.\label{tab:stats}}
\tablewidth{0pt}
\tablehead{
\colhead{} & \colhead{FRB Sample} & \colhead{$n$} & \colhead{$r$} & \colhead{$\log_{10} p$}}
\startdata
\multirow{5}{*}{$\mathrm{DM_{host}}$--$M_\star$}
 & Full sample                                            & 37 & -0.50 & -2.75 \\
 & $9.0 \leq \log_{10}(M_\star/M_\odot) \leq 10.5$        & 23 & -0.03 & -0.05 \\
 & Edge-on hosts ($i > 70^\circ$) excluded                & 33 & -0.57 & -3.27 \\
 & FRBs with $\mathrm{DM_{host}}$ consistent with zero excluded     & 31 & -0.36 & -1.35 \\
 & PRS-hosting FRBs excluded                              & 34 & -0.41 & -1.83 \\
\hline
\multirow{2}{*}{$|\mathrm{RM}|_{\rm host}$--$M_\star$}
 & Full sample                                            & 27 & -0.40 & -1.39 \\
 & PRS-hosting FRBs excluded                              & 24 & -0.01 & -0.01 \\
\enddata
\tablecomments{Pearson correlation coefficient $r$ and the two-sided
probability $p$ of no correlation, computed between
$\log_{10}(M_\star/M_\odot)$ and $\log_{10}\mathrm{DM_{host}}$ or
$\log_{10}|\mathrm{RM}|_{\rm host}$; $n$ is the number of hosts entering
each subset. Both $\mathrm{DM_{host}}$ and $|\mathrm{RM}|_{\rm host}$ are
rest-frame quantities, with
$|\mathrm{RM}|_{\rm host} = |\mathrm{RM_{obs}} - \mathrm{RM_{MW}}|\,(1+z)^2$.
The six hosts with a $\mathrm{DM_{host}}$ consistent with zero
(Table~\ref{tab:frb-sample}) enter at the medians of their
$\mathrm{DM_{host}}$ posteriors, except in the subset that excludes them.
The three PRS-hosting FRBs are identified in Table~\ref{tab:frb-sample}.}
\end{deluxetable*}